\documentclass[12pt]{article}
\usepackage{graphicx}
\def\be{\begin{equation}}
\def\ee{\end{equation}}
\def\bea{\begin{eqnarray}}
\def\eea{\end{eqnarray}}

\def\pe2{p_E^2}

\makeatletter

\@addtoreset{equation}{section}
\makeatother

\begin{document}
\newcommand{\mpl}{M_{\mathrm{Pl}}}
\setlength{\baselineskip}{18pt}
\begin{titlepage}
\begin{flushright}
KOBE-TH-04-05 \\
RIKEN-TH-29 
\end{flushright}

\vspace{1.0cm}

\centerline{{\LARGE\bf An Attempt to Solve the Hierarchy Problem}}
\vspace*{3mm}  
\centerline{{\LARGE\bf  Based on}} 
\vspace*{3mm}  
\centerline{{\LARGE\bf  Gravity-Gauge-Higgs Unification Scenario }} 

\vspace{25mm}

{\Large
\centerline{ K. Hasegawa$^{(a)}$
\footnote{e-mail : kouhei@phys.sci.kobe-u.ac.jp},
C. S. Lim$^{(b)}$
\footnote{e-mail : lim@phys.sci.kobe-u.ac.jp}
and Nobuhito Maru$^{(c)}$ \footnote{e-mail : 
maru@riken.jp}}}
\vspace{1cm}
\centerline{$^{(a)}${\it Graduate School of Science and Technology, 
Kobe University,}}
\centerline{\it Rokkodai, Nada, Kobe 657-8501, Japan}
\centerline{$^{(b)}${\it Department of Physics, Kobe University,
Rokkodai, Nada, Kobe 657-8501, Japan}}
\centerline{$^{(c)}${\it Theoretical Physics Laboratory, RIKEN, 
Wako, Saitama 351-0198, Japan}}

%
%
\vspace{2cm}
\centerline{\large\bf Abstract}
\vspace{0.5cm}
We discuss a possible scenario to solve the hierarchy problem, in which 4-dimensional 
bosonic fields with 
all possible integer spins, graviton, gauge boson and Higgs are unified in a framework of a 
gravity theory with extra dimensions. The Higgs is identified with the extra space 
component of the metric tensor. One-loop quantum effect on the Higgs mass-squared is explicitly calculated 
in a five dimensional gravity theory compactified on $S^1$. 
We obtain a finite calculable Higgs mass-squared without suffering from quadratic divergence, 
by virtue of general coordinate transformation invariance, which is argued to be guaranteed by 
the summation over all Kaluza-Klein modes running in the loop diagrams.  
\end{titlepage}

%
%
\section{Introduction} 
One of the long standing issues in the particle physics 
is the hierarchy problem: 
the problem of how to maintain the hierarchy 
between two mass scales of the theory, 
$M_{W}$ and $M_{GUT}$ or $M_{pl}$, which differ by many orders of magnitude.  
When the standard model is regarded as a ``low-energy" effective theory 
with a physical cutoff $\Lambda \sim M_{GUT}, \ M_{pl}$, 
the Higgs mass-squared $m_{H}^{2}$ seems to get 
a quantum correction $\sim \Lambda^{2}$, 
invalidating the hierarchy tuned at the classical level: 
the problem of ``quadratic divergence". 

Since the hierarchy problem has been historically 
playing a central role in the development of the physics 
beyond the standard model, 
we believe that attempts to exhaust the possibilities 
to solve the problem is quite important. 
In particular, 
to exploit the mechanisms to cancel the quadratic divergence 
will be helpful in getting insight into the hidden symmetries 
in the physics beyond the standard model. 
 
Conventional wisdom in four dimensional space-time 
to solve the problem of the quadratic divergence is 
to rely on the supersymmetry. 
Recently it has been realized that 
alternative scenarios to solve the hierarchy problem are possible 
once we extend our space-time {\cite{Arkani, RS1, HIL}}. 
The authors of {\cite{Arkani, RS1}} adopted higher dimensional gravity theories, and aimed to solve the hierarchy problem 
between $M_{pl}$ and $M_{W}$, invoking to large extra dimension {\cite{Arkani}}, or to the 
``warp factor" appearing in the metric of non-factorizable AdS$_5$ 
space-time with 3-branes{\cite{RS1}}, 
though the hierarchy was discussed at the classical level. 
The approach taken in {\cite{HIL}} is a bit different: 
it deals with higher dimensional gauge theories 
where the Higgs field is identified with 
the extra space component of gauge field, 
and the main concern was the problem at the quantum level, 
i.e. the problem of quadratic divergence.  
In the scenario, the gauge boson and Higgs scalar with different spins (from 4-dimensional (4D) point of view) 
are unified as a gauge boson in higher dimensional space-time; 
``Gauge-Higgs unification" is realized. 
The finiteness of the Higgs mass, without suffering from the quadratic divergence, is guaranteed 
by the higher dimensional local gauge symmetry, 
whose transformation is due to a parameter 
depending on the extra-space coordinates. 

It has been argued {\cite{HIL}} that 
in the calculation of the Higgs mass-squared, 
the summation over all Kaluza-Klein (K-K) modes 
in the intermediate state of the loop diagram is inevitable 
in order to preserve the higher dimensional gauge symmetry, necessary to get the finite Higgs mass.  
This is because the momentum cutoff generally spoils the gauge invariance, 
while the K-K mode corresponds to the extra-space component of the momentum. 
In fact, it was demonstrated by explicit calculation that 
the K-K mode sum provides a finite calculable Higgs mass.  
The idea itself to identify the extra space component of gauge field with the Higgs field is not 
new {\cite{Manton}}. In particular, dynamical gauge symmetry breaking 
due to the vacuum expectation value (VEV) of the extra space component 
 has been argued to be possible in the non-Abelian theories {\cite{Hosotani}}. 

Attempts to construct realistic theories beyond the standard model based on 
the Gauge-Higgs unification scenario have been already made, 
utilizing orbifolding of extra spaces with non-trivial $Z_{2}$-parity 
assignment for the fields in order to break gauge symmetries \cite{Kawamura}, 
with minimal SU(3) gauge symmetry \cite{Kubo} or more elaborate gauge symmetries and/or 
more extra spaces \cite{Murayama}. 
It is worth while noticing that another interesting scenario to get stabilized Higgs mass, 
``dimensional de-construction" \cite{Georgi}, may be understood as a sort of 5D gauge theory where the extra 5-th dimension is ``latticized". In fact, we can check that the effective potential of the Higgs scalar obtained in the scenario just coincides with that in the Gauge-Higgs Unification in the limit of 
$N \ \to \ \infty$ ($N$: the number of lattice sites). There is also an interesting claim that 
the Gauge-Higgs unification scenario may have an important cosmological implication, stabilizing 
the inflaton potential under the quantum correction \cite{AR}.

It is interesting to note that in both of the supersymmetry 
and the Gauge-Higgs unification scenarios, 
the 4D Poincar$\acute{\mbox{e}}$ symmetry is somehow enlarged. 
In the case of supersymmetry, 
the Poincar$\acute{\mbox{e}}$  symmetry is extended to that of superspace. 
Accordingly, 4D fields with different spins $(H, \ \psi_{H})$, 
with $\psi_{H}$ being Higgsino, are unified in a super-multiplet. 
The smallness of the Higgs mass $m_{H}$ is then related, 
via supersymmetry, to that of $m_{\psi_{H}}$, 
which in turn is attributed to 
the chiral symmetry of $\psi_{H}$ sector.  
Similarly, in the case of Gauge-Higgs unification, 
the Poincar$\acute{\mbox{e}}$ symmetry is extended to that of 5D space-time, 
and 4D fields with different spins $(H, \ A_{\mu})$, with $A_{\mu}$  
being a gauge boson, are unified in a form of the higher dimensional gauge boson $A_{M}$. 
The smallness of the Higgs mass $m_{H}$ is then related, 
via the higher dimensional Poincar$\acute{\mbox{e}}$ symmetry, 
to the vanishing mass of $A_{\mu}$, which is attributed to 
the ordinary 4D gauge symmetry of $A_{\mu}$ sector.  

As the matter of fact,  this Poincar$\acute{\mbox{e}}$ symmetry is ``softly" broken 
by the presence of the compactification scale $1/R$ 
($R$ being a generic size of the extra-space, such as the radius of sphere), 
thus leading to the finite mass $m_{H}$, 
roughly proportional to $1/R$ for $S^{1}$ extra-space, for instance.  
This finite mass may also be understood as the consequence of 
the appearance of non-local gauge invariant operator, i.e. 
a non-trivial Wilson loop along $S^{1}$;  
$W = P \ e^{ig \oint A_{y} \ dy}$ ($\grave{\mbox{a}}$ la A-B effect). 

In the Gauge-Higgs unification scenario 
the bosonic fields with spins $s = 1$ and $0$ are unified. 
Then it may be a natural question to ask, 
whether a unification of all bosonic states with the highest spin $2$ is ever possible. 

In this paper, we investigate this possibility. 
Namely, we extend the Gauge-Higgs unification scenario, 
and propose a mechanism to solve the hierarchy problem  
in the framework of ``Gravity-Gauge-Higgs unification", 
where all known bosonic particles with different spins, 
i.e. graviton, gauge boson and Higgs, 
are unified in the scheme of higher dimensional gravity theory. 
We thus make the extended Poincar$\acute{\mbox{e}}$ symmetry local. 
We will discuss, as a prototype model, 5D gravity theory, 
i.e. the original Kaluza-Klein theory, 
though we also introduce a matter field to make our argument simple and 
transparent. 
The Higgs field is identified with the extra space component 
$g_{55}$ of the metric tensor, whose mass exactly vanishes at the classical level.  
Then the stability of the Higgs mass under the quantum correction 
is naturally guaranteed by the local symmetry of the gravity theory, 
i.e. by the general coordinate transformation invariance, 
instead of the local gauge invariance in the Gauge-Higgs unification scenario.  
  
Our main purpose in this paper is to demonstrate that the mechanism works. Namely, 
we will show by an explicit calculation, 
that, by virtue of the summation over all Kaluza-Klein (K-K) modes 
in the intermediate state of the loop diagram, 
we get a finite calculable quantum correction to the Higgs mass, without suffering from 
a quadratic divergence.  
We first calculate the quantum correction due to the introduced 5D scalar matter field. 
Then the obtained result is readily generalized to the quantum corrections from 
a variety of fields with different spins, including graviton itself.    

\section{A prototype model} 
For the purpose to illustrate the mechanism of 
the cancellation of the quadratic divergence, 
in this paper we discuss a prototype model: 
5D gravity theory described by a metric tensor 
$g_{MN} \ (M = \mu \ (0,1,2,3) \ \mbox{or}~5)$. 
We identify (the K-K zero-mode of) $g_{55}$ with our Higgs field; 
to be precise the K-K zero-mode of $g_{55}$ is written as 
$g_{55} = - e^{\phi}, \ \phi = \phi_{0} + h $, 
where $\phi_{0}$ is the VEV of 
the $\phi$ field and $h$ corresponds to the Higgs field.  
(Actually, we will see below that, as the field $h$ is dimensionless, 
the field $H = \frac{\sqrt{6}}{4}\frac{h}{\sqrt{\kappa}}, \ \ \ (\kappa \equiv 8\pi G)$ 
should be identified with the physical Higgs field.) 
The 5D space-time coordinates are 
\be
\label{5Dspacetime} 
(x^{\mu}, y) \ \ \ (0 \le y < 2\pi R).  
\ee  
where the extra space with the coordinate $y$ 
is assumed to be compactified on $S^{1}$, 
whose ``physical" radius $\hat{R}$ is given in terms of the zero-mode of $g_{55}$ 
as\footnote{In this paper, 
we take the metric convention: 
$\eta_{MN} = {\rm diag}(+1,-1,-1,-1,-1)$. 
Ricci tensor ${\cal R}_{MN}$ is defined as 
${\cal R}_{MN} \equiv {\cal R}^P_{MNP} \equiv g^{PQ}{\cal R}_{PMNQ}$. 
} 
\be
\label{radius} 
2\pi \hat{R} = \int_{0}^{2\pi R}  \ \sqrt{- g_{55}} \ dy = 
2\pi R~e^{\frac{\phi}{2}}. 
\ee 
The size of the $S^{1}$ is fixed to be $\hat{R}_{0} = R \ e^{\frac{\phi_{0}}{2}}$, 
once  the VEV $\phi_{0}$ is determined 
by the minimization of the radiatively induced effective potential 
of $\phi$, $V_{\rm{eff}} (\phi)$: ``spontaneous  compactification".  

As the matter field 
it may be realistic to introduce some fermions. 
In the present paper, however, 
we introduce a 5D bulk scalar field $\Phi$ for the sake of 
the computational simplicity of the radiative correction 
to the Higgs mass $m_{H}$. 
Note that the cancellation mechanism of the quadratic divergence 
is based on the general coordinate invariance 
and no matter what kind of matter field we choose 
we should be able to obtain a finite $m_{H}$ 
as long as the K-K mode sum is kept. 
After getting the finite radiative correction to $m_{H}$ 
due to the scalar field  $\Phi$, the obtained result turns out to be easily generalized 
for the contributions from other fields with various spins, such as fermion and the graviton itself, 
just by counting the physical degrees of freedom of polarization.    

The action we consider is given by
\be 
S = S_{g} + S_{s}, 
\ee 
where 
\bea
\label{action:g} 
S_{g} &=& \frac{1}{16\pi G_{5}} \ \int \ d^{4}x dy \ \sqrt{g} \ {\cal R},  \\ 
\label{action:s}
S_{s} &=& \int \ d^{4}x dy \ \sqrt{g} \ \frac{1}{2} g^{MN} \ 
(\partial_{M} \Phi) (\partial_{N} \Phi), 
\eea 
where $G_{5}$ is the 5D gravitational constant, 
and $g$ and ${\cal R}$ should be calculated from $g_{MN}$. 

The stability of the Higgs mass $m_H$ under the quantum correction is guaranteed 
by the fact that the Higgs field transforms inhomogeneously 
under the general coordinate transformation, 
whose transformation parameter is $y$-dependent. 
To see the transformation property we write an infinitesimal line element $ds$ as 
\be
\label{ds2}  
ds^{2} = g_{\mu \nu} dx^{\mu} dx^{\nu} 
       - e^{\phi} \ (dy + A_{\mu} dx^{\mu})^{2}.  
\ee 
Now a special sort of infinitesimal general coordinate transformation 
\bea
\label{gct:x} 
x^{\mu} &\to& x^{\prime \mu} = x^{\mu}, \\ 
\label{gct:y}
y       &\to& y^{\prime} = y + \alpha (x^{\mu}, y), 
\eea  
with the transformation parameter $\alpha$, 
causes the transformation of fields  
\bea 
\label{trf:gmn}
g_{\mu \nu} &\to& g^{\prime}_{\mu \nu} = g_{\mu \nu}, \\ 
\label{trf:am}
A_{\mu}     &\to& A^{\prime}_{\mu} = (1 + \frac{\partial \alpha}{\partial y}) 
A_{\mu} - \frac{\partial \alpha}{\partial x^{\mu}},  \\ 
\label{trf:phi}
\phi       &\to& \phi^{\prime} = \phi - 2 \frac{\partial \alpha}{\partial y}.  
\eea  
If we let $\alpha$ dependent only on $x^{\mu}$, 
we obtain an ordinary U(1) gauge transformation for $A_{\mu}$,  
which is the original idea of Kaluza and Klein. 
If, instead, we let $\alpha$ dependent only on $y$, 
we obtain the inhomogeneous transformation for the field $\phi$, 
together with a suitable scale transformation of $A_{\mu}$. 
According to (\ref{ds2}), we write the 5D metric tensor in the form 
\be 
\label{5Dmetric}
g_{MN} = 
\pmatrix{
g_{\mu \nu} - e^{\phi} A_{\mu} A_{\nu} & - e^{\phi} A_{\mu} \cr
         - e^{\phi} A_{\nu}  & - e^{\phi} 
}. 
\ee 

In order to show that the effective low energy theory of $g_{MN}$ 
is described by the unified system of 
4D graviton $g_{\mu \nu}$, gauge boson $A_{\mu}$ and Higgs $\phi$, 
we write down $S_{g}$ in terms of zero-modes of these fields, 
ignoring the $y$-dependence of these fields: 
\be
\label{0modes} 
S_{g} =  \frac{2\pi R}{16\pi G_{5}} \ \int \ d^{4}x \ \sqrt{- g_{4}} \ 
e^{\frac{\phi}{2}}
 \ \left({\cal R}^{(4)} - \frac{1}{4} e^{\phi} F_{\mu \nu} F^{\mu \nu} \right),  
\ee  
where $g_{4}$ and ${\cal R}^{(4)}$ are calculated from $g_{\mu \nu}$ alone and 
$F_{\mu \nu} = \partial_{\mu} A_{\nu} - \partial_{\nu} A_{\mu}$ \cite{WR}. 
This result, at the first glance, seems to mean that 
the Higgs field does not appear as a dynamical variable in the low energy theory. 
On the other hand, however, 
it is well-known that once an appropriate gauge conditions, 
i.e. harmonic condition and traceless condition, 
are imposed on the metric $g_{MN}$,  the each component field of the metric tensor 
is separated in the weak field approximation, 
and the resultant equation of motion is just 
Klein-Gordon type equation for the graviton; we do expect to have $\phi$ as a dynamical 
variable. In fact, we can show that, by use of the weak field approximation,  
$g_{\mu \nu} = \eta_{\mu \nu} + h_{\mu \nu}, e^{\phi} = e^{\phi_{0}}(1 + h)$ 
with $|h_{\mu \nu}|, \ |A_{\mu}|, \ |h| \ll 1$, the action, when it is collaborated with the harmonic 
and traceless conditions for zero-modes, 
$\partial^{\mu} h_{\mu \nu} = 0, \ \partial^{\mu} A_{\mu} = 0, \ \eta^{\mu \nu} h_{\mu \nu} - h = 0$,  
is written as 
\be 
\label{wfpt}
S_{g} =  \frac{1}{16\pi G} \ \int \ d^{4}x \  
\left\{\frac{1}{4} (\partial_{\mu} h_{\alpha \beta})
(\partial^{\mu} h^{\alpha \beta}) 
- \frac{e^{\phi_{0}}}{2} (\partial_{\mu} A_{\nu})(\partial^{\mu} A^{\nu}) 
+ \frac{1}{4} (\partial_{\mu} h)(\partial^{\mu} h)\right\},   
\ee 
where the 4D gravitational constant is given by
$G \equiv G_{5}/(e^{\frac{\phi_{0}}{2}} 2\pi R)$. 
The 4D metric $h_{\alpha \beta}$, however, still partially contains $h$ as its traceful part, 
because of the 5-dimensional traceless condition $\eta^{\alpha \beta} h_{\alpha \beta} - h = 0$. 
Hence, it is necessary to separate $h$ from $h_{\alpha \beta}$, so that the remaining 
$h_{\alpha \beta}$ really stands for the 4D graviton. For such purpose, we consider the 
physical degree of freedom of $h_{\alpha \beta}$, i.e. the metric with only 
transverse polarization, which we denote by 
$h^{t}_{\tilde{\alpha} \tilde{\beta}}$ \ ($\tilde{\alpha}, \tilde{\beta} = 2, 3$  
for the momentum in the $x$-direction, for instance). $h^{t}_{\tilde{\alpha} \tilde{\beta}}$ is 
obtainable by a suitable general coordinate transformation, consistent with the 
harmonic and traceless conditions. 
$h^{t}_{\tilde{\alpha} \tilde{\beta}}$ can be decomposed into a traceless part 
$\hat{h}_{\tilde{\alpha} \tilde{\beta}}$ and a traceful part proportional to $h$: 
$h^{t}_{\tilde{\alpha} \tilde{\beta}} = \hat{h}_{\tilde{\alpha} \tilde{\beta}} + \frac{1}{2} 
\eta_{\tilde{\alpha} \tilde{\beta}} h \ 
(\eta^{\tilde{\alpha} \tilde{\beta}} \hat{h}_{\tilde{\alpha} \tilde{\beta}} = 0, \ 
\eta^{\tilde{\alpha} \tilde{\beta}} h^{t}_{\tilde{\alpha} \tilde{\beta}} - h = 0 )$. 
Substituting this decomposition of $h^{t}_{\tilde{\alpha} \tilde{\beta}}$ for $h_{\alpha \beta}$ in 
(\ref{wfpt}), and changing $\tilde{\alpha} \tilde{\beta}$ into $\alpha \beta$ 
in order to recover the 4D Lorentz covariance, we get 
\be 
\label{physical} 
S_{g} =  \frac{1}{16\pi G} \ \int \ d^{4}x \  
\left\{\frac{1}{4} (\partial_{\mu} \hat{h}_{\alpha \beta})
(\partial^{\mu} \hat{h}^{\alpha \beta}) 
- \frac{e^{\phi_{0}}}{2} (\partial_{\mu} A_{\nu})(\partial^{\mu} A^{\nu}) 
+ \frac{3}{8} (\partial_{\mu} h)(\partial^{\mu} h)\right\},   
\ee  
where $\hat{h}_{\alpha \beta}$ satisfying 4D harmonic and traceless conditions, 
$\partial^{\alpha} \hat{h}_{\alpha \beta} = 0, \ \eta^{\alpha \beta} \hat{h}_{\alpha \beta} = 0$, 
should be identified with the 4D graviton. 
This implies the Higgs field $H$ with correct mass dimension 
and canonical kinetic term in 4D space-time should be identified as 
\be
\label{canonical} 
H = \frac{\sqrt{6}}{4}\frac{h}{\sqrt{\kappa}}, \ \ \ (\kappa \equiv 8\pi G). 
\ee

It is trivial that Einstein equation has the following constant background solution 
for the metric tensor (with the vanishing matter field $\Phi$ being understood):
\be
\label{classicalsol}
g_{\mu\nu} =\eta_{\mu\nu}, \quad A_\mu = 0. \quad \phi = {\rm constant}, 
\ee
which implies that the background space-time is $M^4 \times S^1$ 
with the radius of $S^1$, $\hat{R}$ being given by (\ref{radius}). 
At the classical level, 
the value of the constant $\phi$ or the radius of $S^1$ is not fixed. 
At the quantum level, however, 
we will have nontrivial effective potential $V_{\rm{eff}}$ with respect to $\phi$, 
which in principle determines the size of $S^1$. 

\section{Explicit diagrammatic calculation of the Higgs mass}
In this section, we will explicitly calculate the two point function of $h$ 
by use of Feynman diagrams. We wish to demonstrate that a finite calculable 
mass for $h$ is obtained 
by virtue of the summation over all K-K modes in the intermediate states 
in the loop diagram. 
For simplicity, 
here we calculate the quantum effects due to a bulk scalar field 
$\Phi$ by use of $S_s$. 
Such a thing is justified since the invariance under the general coordinate 
transformation holds separately in each of $S_g$ and $S_s$, 
and $\Phi$ contribution alone should provide a finite mass-squared 
for $h$, $m_{h}^{2}$. 
Later we will also discuss the contribution from $S_g$, 
namely the quantum effect due to the 5D graviton (or self-interactions 
among $\phi, g_{\mu\nu}$ and $A_\mu$).

As we are familiar with the Feynman rule in 4D space-time, let us derive 4D action obtained from $S_s$ 
by performing $y$-integral. 
For such purpose, we make mode expansion of $\Phi(x,y)$ 
\be 
\label{expand}
\Phi(x,y) \equiv \sum_n \frac{1}{\sqrt{2\pi \hat{R}}} \Phi_n(x) e^{iny/R}, \ \ \ 
(\Phi_n(x) = \Phi_{-n}(x)^{\ast}),   
\ee 
with the ortho-normality condition 
\be  
\label{orthonormal}
\int^{2\pi R}_0 dy \sqrt{-g_{55}}
\left( \frac{1}{\sqrt{2\pi \hat{R}}} e^{imy/R} \right) 
\left( \frac{1}{\sqrt{2\pi \hat{R}}} e^{iny/R} \right) = \delta_{n,-m}.   
\ee  
Then, substituting the mode expansion into (\ref{action:s}), the $y$-integral yields  
\bea
\label{4Daction:s}
S_s = \int d^4x \sum_n \frac{1}{2}
\Phi_n(x) 
\left\{
-\eta^{\mu\nu}\partial_\mu \partial_\nu 
- \left( \frac{n}{\hat{R}} \right)^{2}
\right\}
\Phi_n(x),
\eea
where, since we are interested in the mass-squared term of $h$ field, we have assumed  
that the metric tensor takes its constant background (\ref{classicalsol}). 
(For a given $n > 0$ we have a complex field $\Phi_n(x) = \Phi_{-n}(x)^{\ast}$, which can be 
decomposed into real and imaginary parts: $\Phi_n(x) = \frac{\mbox{Re} \Phi_{n} + i \mbox{Im} \Phi_{n}}
{\sqrt{2}}$. These real and imaginary parts are rewritten in (\ref{4Daction:s}) as 
$\Phi_{n}$ and $\Phi_{-n}$, which should be regarded as two independent real fields. 
The zero-mode $\Phi_{0}(x)$ is real field from the beginning.)     

Note that the 4D action (\ref{4Daction:s}) has canonical 4D kinetic terms   
without a factor $\sqrt{-g_{55}}=e^{\phi/2}$ and 
we can obtain 4D Feynman rules from this action. 
To do this, we expand $\phi$ in the $\hat{R} = R \ e^{\frac{\phi}{2}}$ around 
the VEV $\phi_0$, 
\be
\label{fluctuation}
\phi = \phi_0 + h,
\ee
where $h$ corresponds to the Higgs $H = \frac{\sqrt{6}}{4}\frac{h}{\sqrt{\kappa}}$.  
Thus, up to ${\cal O}(h^2)$, 
(\ref{4Daction:s}) can be expanded as $(\hat{R}_{0} \equiv R \ e^{\frac{\phi_{0}}{2}})$ 
\bea
S_s &=& S_s^{({\rm free})} + S_s^{({\rm int})}, \\
\label{4Dscalar:free}
S_s^{({\rm free})} &=& \int d^4x \sum_n \frac{1}{2} 
\Phi_n(x)
\left\{
-\eta^{\mu\nu}\partial_\mu \partial_\nu 
- \left( \frac{n}{\hat{R}_0} \right)^{2}
\right\}
\Phi_n(x), \\
\label{4Dscalar:int}
S_s^{({\rm int})} &=& -\frac{1}{2} \int d^4x \sum_n 
\left( -h + \frac{1}{2} h^2 \right) \left( \frac{n}{\hat{R}_0} \right)^2 
\Phi_{n}(x)^2. 
\eea
The relevant Feynman rules are read off from (\ref{4Dscalar:free}) and (\ref{4Dscalar:int}) as listed in Fig. 1.

\begin{figure}[ht]
\begin{center}
\includegraphics[width=15cm]{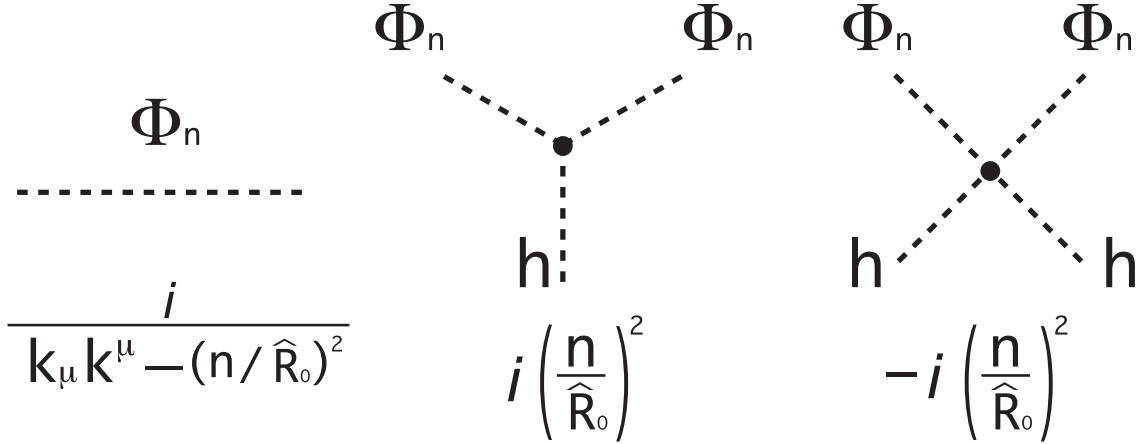}
\caption{Feynman rules relevant for the calculation of $m_{h}^{2}$.}
\label{rules}
\end{center}
\end{figure}

One important remark is that the finiteness of $m^2_h$ is guaranteed 
in the 5D point of view, keeping all K-K modes. 
Thus, in addition to the ordinary 4D Feynman rule, 
some care should be taken when we perform a K-K mode summation $\sum_n$. 
Namely, instead of a simple summation $\sum_n$, 
we should adopt
\be
\frac{1}{2\pi \hat{R}}\sum_n,
\ee
which reduces to $\int \frac{dk_y}{2\pi}$ 
and provides a correct Feynman rule for a non-compact 5D space-time   
in the ``de-compactification" limit $\hat{R} \to \infty$.  
Or together with 4D momentum integral,
\be
\label{remark}
\frac{1}{2\pi \hat{R}}\sum_n \int d^4 k
\ee
should be regarded as the ``trace" in the 5D phase space, 
since in the ordinary Fourier expansion we have Fourier modes 
for every $\frac{1}{2\pi \hat{R}}$. This may be also understood as follows. 
What we are interested in is the radiatively induced $\frac{1}{2} m_{h}^{2} h^{2}$ term in the 5D 
effective potential $V_{\rm{eff}}^{(5D)}$, which is related to 4D effective potential 
$V_{\rm{eff}}^{(4D)}$ as $V_{\rm{eff}}^{(4D)} = \int_{0}^{2\pi R} dy \sqrt{- g_{55}} 
V_{\rm{eff}}^{(5D)} 
= 2\pi \hat{R} V_{\rm{eff}}^{(5D)}$. Thus to get $V_{\rm{eff}}^{(5D)}$ the 4D result should be multiplied 
by $1/(2\pi \hat{R})$.  
With the prescription (\ref{remark}) for the loop integral, 
we may calculate the effective $h^2$ operator 
just according to the Feynman rules in Fig.1.

We should pay attention to the fact that not only the diagrams with two external $h$ lines, but also 
bubble and tadpole diagrams should be evaluated,  since the factor  
\be
\frac{1}{\hat{R}} = \frac{1}{\hat{R}_0}e^{-h/2} 
\simeq \frac{1}{\hat{R}_0}(1 - \frac{h}{2} + \frac{h^2}{8})
\ee
in (\ref{remark}) is $h$-dependent. 

Thus, the relevant diagrams we should compute are those shown in Fig.2.  

\begin{figure}[ht]
\begin{center}
\includegraphics[width=15cm]{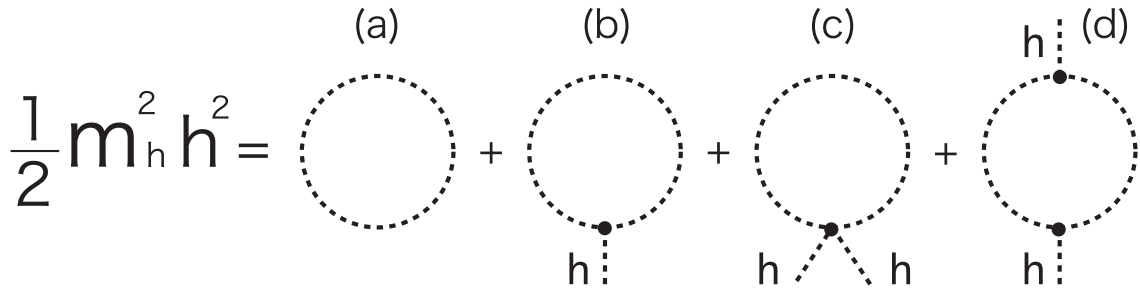}
\caption{Diagrams relevant for the 1-loop corrections to the $m_{h}^{2}$.}
\label{allgraph}
\end{center}
\end{figure}

The contribution of each diagram to $h^2$ operator is
\bea
\label{graph:a}
(a)&& \frac{1}{2\pi \hat{R}} \sum_n \int \frac{d^4k}{(2\pi)^4} 
\left( -\frac{i}{2} \right) 
\ln \left[ -k_\mu k^\mu + \left( \frac{n}{\hat{R}_0} \right)^2 \right], 
\nonumber \\
\label{graph:a1}
&& \to \frac{1}{2\pi \hat{R}_0} \sum_n \int \frac{d^4k}{(2\pi)^4} 
\left( -\frac{i}{2} \right) \frac{1}{8}
\ln \left[ -k_\mu k^\mu + \left( \frac{n}{\hat{R}_0} \right)^2 \right]h^2, \\
\label{graph:b}
(b)&& \frac{1}{2\pi \hat{R}} \sum_n \int \frac{d^4k}{(2\pi)^4} 
\left( \frac{i}{2} \right) i\left( \frac{n}{\hat{R}_0} \right)^2
\frac{i}{ k_\mu k^\mu - \left( \frac{n}{\hat{R}_0} \right)^2 }h, 
\nonumber \\
\label{graph:b1}
&& \to \frac{1}{2\pi \hat{R}_0} \sum_n \int \frac{d^4k}{(2\pi)^4} 
\left( \frac{i}{4} \right) 
\frac{ \left( \frac{n}{\hat{R}_0} \right)^2}
{ k_\mu k^\mu - \left( \frac{n}{\hat{R}_0} \right)^2 }h^2, \\
\label{graph:c}
(c)&& \frac{1}{2\pi \hat{R}} \sum_n \int \frac{d^4k}{(2\pi)^4} 
\left( \frac{i}{4} \right) (-i)\left( \frac{n}{\hat{R}_0} \right)^2
\frac{i}{ k_\mu k^\mu - \left( \frac{n}{\hat{R}_0} \right)^2 }h^2, 
\nonumber \\
\label{graph:c1}
&& \to \frac{1}{2\pi \hat{R}_0} \sum_n \int \frac{d^4k}{(2\pi)^4} 
\left( \frac{i}{4} \right) \frac{ \left( \frac{n}{\hat{R}_0} \right)^2}
{ k_\mu k^\mu - \left( \frac{n}{\hat{R}_0} \right)^2 }h^2, \\
\label{graph:d}
(d)&& \frac{1}{2\pi \hat{R}} \sum_n \int \frac{d^4k}{(2\pi)^4} 
\left( \frac{i}{4} \right) 
\left\{ i\left( \frac{n}{\hat{R}_0} \right)^2 \right\}^2
\left\{ \frac{i}{ k_\mu k^\mu - \left( \frac{n}{\hat{R}_0} \right)^2 } 
\right\}^2 h^2, 
\nonumber \\
\label{graph:d1}
&& \to \frac{1}{2\pi \hat{R}_0} \sum_n \int \frac{d^4k}{(2\pi)^4} 
\left( \frac{i}{4} \right) \frac{ \left( \frac{n}{\hat{R}_0} \right)^4 }
{ \left\{ k_\mu k^\mu - \left( \frac{n}{\hat{R}_0} \right)^2 \right\}^2} h^2. 
\eea
Combining these results, we obtain the induced mass squared $m^2_h$
\bea
\label{scalarmass:g}
m^2_h &=& \left( -\frac{i}{2} \right) \left( \frac{1}{2\pi \hat{R}_0} \right)
\sum_n \int \frac{d^4k}{(2\pi)^4} \nonumber \\
&& \times \left\{ \frac{1}{4} 
\ln \left[ -k_\mu k^\mu + \left( \frac{n}{\hat{R}_0} \right)^2 \right] 
-2\frac{\left( \frac{n}{\hat{R}_0} \right)^2}
{k_\mu k^\mu - \left( \frac{n}{\hat{R}_0} \right)^2}
- \frac{\left( \frac{n}{\hat{R}_0} \right)^4}
{\left\{ k_\mu k^\mu - \left( \frac{n}{\hat{R}_0} \right)^2 \right\}^2}
\right\}. 
\eea

We expect that this result for $m^2_h$ should be finite (not UV divergent).  
To be more precise, the $m^2_h$ should vanish in the limit of ``de-compactification" 
$\hat{R} \to \infty$, since the local operator $h^{2}$ is forbidden by the invariance under 
the general coordinate transformation, under which $h$ transforms inhomogeneously (recall 
(\ref{trf:phi}) and (\ref{fluctuation})). When $\hat{R}$ becomes finite, we expect to have a non-vanishing $m^2_h$, but 
it should be still finite. This is because the difference between the finite and 
the infinite radius cases appears in the infrared region of 5-th momentum $k_{y}$. 
Therefore, UV divergence is insensitive to the finiteness of the radius, and as long as 
we have vanishing $m^2_h$ for the de-compactification limit, we will have, at most, some finite 
$m^2_h$ for the compactified space. 

To show this fact explicitly we apply the  de-compactification limit $\hat{R} \to \infty$ 
for the result of $m^2_h$ (\ref{scalarmass:g}) 
\be 
\label{masspot1}
m^2_h 
\to 
-\frac{i}{2} 
\int \frac{d^4k dk_y}{(2\pi)^5} 
\left\{
\frac{1}{4}{\rm ln}\left[ -k_\mu k^\mu + k_y^2 \right] 
+ \frac{2k_y^2}{-k_\mu k^\mu + k_y^2} 
- \frac{k_y^4}{\left[-k_\mu k^\mu + k_y^2 \right]^2}
\right\}. 
\ee 
The obtained result can be written as
\bea
\label{masspot2}
m^2_h &=& \left. 
\left[ \frac{1}{4} I(\alpha) +2 \frac{d}{d \alpha^2} I(\alpha) 
+ \left( \frac{d}{d \alpha^2} \right)^2 I(\alpha)\right] 
\right|_{\alpha^2=1} \\
&=& \left. 
\left[ \frac{1}{4\alpha} -\frac{1}{\alpha^3} +\frac{3}{4\alpha^5} 
\right] \right|_{\alpha=1}\tilde{I} 
= \left( \frac{1}{4} -1 + \frac{3}{4} \right) \tilde{I} = 0,
\eea
in terms of a useful integral 
\bea
\label{integral}
I(\alpha) &\equiv& -\frac{i}{2} \int \frac{d^4k dk_y}{(2\pi)^5} 
\ln [- k_\mu k^\mu + \alpha^2 k_y^2] \\
&=& -\frac{i}{2\alpha} \int \frac{d^4k d\tilde{k}_y}{(2\pi)^5} 
\ln [- k_\mu k^\mu + \tilde{k}_y^2] 
\equiv \frac{1}{\alpha} \tilde{I}.  
\eea 
Thus we have confirmed that $m_{h}^{2}$ does disappear in the limit.  

The finite $m_{h}^{2}$ for the finite radius case is given by
\be
\label{finitemass}
m^2_h = \left. \left[ \frac{1}{4} 
+ \frac{1}{\alpha} \frac{d}{d \alpha} 
+ \frac{1}{4}\left( \frac{1}{\alpha} \frac{d}{d\alpha} \right)^2 \right]
\hat{I}(\alpha, \hat{R}_{0}) \right|_{\alpha=1},  
\ee
where an integral similar to (\ref{integral}) for the finite radius 
is defined as \cite{Hosotani2} 
\bea
\label{integral1}
\hat{I}(\alpha,\hat{R}_{0}) &\equiv& -\frac{i}{2}\frac{1}{2\pi \hat{R}_{0}} 
\sum_n \int \frac{d^4k}{(2\pi)^4} \ln \left[ -k_\mu k^\mu 
+ \alpha^2 \left( \frac{n}{\hat{R}_{0}} \right)^2 \right], \\
\label{finite}
&=& \frac{1}{\alpha} 
\left[ \tilde{I}    
- \frac{3\alpha^5 \zeta(5)}{128 \pi^7 \hat{R}_{0}^{5}}
\right],
\eea
where
\be 
\zeta(5) = \sum_{n=1}^\infty \frac{1}{n^5}. 
\ee 
Substituting (\ref{finite}) into  (\ref{finitemass}), 
we obtain the finite scalar mass 
\be
\label{finitemass2} 
m_h^2 = -\frac{75\zeta(5)}{512 \pi^7 \hat{R}_0^5}.  
\ee
where one can explicitly see that the UV divergent constant $\tilde{I}$ disappears, as we expected.  
Finally, the finite mass for the Higgs field $H$ 
with correct mass dimension is obtained as\footnote{First factor is 
required to obtain the 4D effective potential from the 5D one and the 
second factor comes from (\ref{canonical}).}
\be 
\label{higgsmass} 
m_H^2 = 2 \pi \hat{R}_0 \times \frac{64}{3} \pi G \times m_h^2 
= -\frac{25\zeta(5)}{4\pi^5 M_{pl}^2 \hat{R}_0^4}, 
\ee
where $M_{pl}$ is a 4D Planck mass. The finite mass should be understood to be due to some non-local 
(global or infrared) effect and is general coordinate transformation invariant since 
the mass depends only on $\hat{R}_0$.

It is important to note that if we retain only zero mode ($n=0$) 
in (\ref{scalarmass:g}), we have a UV divergent result:
\be
\label{0mode}
m^2_h = -\frac{i}{2}\frac{1}{2\pi \hat{R}_{0}} \int 
\frac{d^4k}{(2\pi)^4} \frac{1}{4}\ln(-k_\mu k^\mu) = \infty. 
\ee
This is consistent with the fact that 5D general coordinate 
transformation invariance guaranteed by the K-K mode sum ensures the finiteness of the Higgs mass. 

Though we have assumed that $\phi$ develops a VEV $\phi_{0}$, (\ref{fluctuation}), 
we actually find that the linear term of $h$ is also induced radiatively 
with a finite but non-vanishing coefficient in this prototype model. 
This signals the instability of the effective potential $V_{{\rm eff}}$ 
\cite{AC}. As we discuss below, the stabilization of $V_{{\rm eff}}$ 
and therefore the stabilization of the radius of $S^{1}$ can be realized 
by adding, e.g., massive bulk matter fields. The formula (\ref{finitemass2}) 
will be modified by the extension of the model. 
Our purpose here, however, is to demonstrate how the finite $m_{H}$ is 
obtained by the K-K mode sum and the interplay between different types of 
Feynman diagrams. Even if such additional fields are included, we still 
get a finite $m_{H}$, since the general coordinate transformation invariance holds 
in each sector of the fields $h$ couples with. Hence, once a realistic model is 
provided, we can readily calculate the finite $m_{H}$ according to the prescription shown 
right above. 

\section{Effective potential approach} 
The Higgs mass-squared obtained above may be more systematically obtained by considering 
effective potential of $\phi$, $V_{{\rm eff}}(\phi)$ induced by the quantum effect of $\Phi$.  
Although the calculations and results here are not new \cite{AC}, 
the previous works are not focused on the Higgs mass and the hierarchy problem.  
Therefore, we rewrite the known results so that it becomes relevant for the Higgs mass.  
The 1-loop effective potential we calculate is 
\be
\label{effpot}
V_{{\rm eff}} = -\frac{i}{2} \left( \frac{1}{2\pi \hat{R}} \right) 
\sum_n \int \frac{d^4k}{(2\pi)^4}~{\rm ln}
\left[-k_\mu k^\mu + \left( \frac{n}{\hat{R}} \right)^2 \right]. 
\ee
Note that the factor $\left( \frac{1}{2\pi \hat{R}} \right)$ is required 
in summing K-K modes in order to have a natural correspondence of 
the 5-th momentum integral in the $\hat{R} \to \infty$ limit. 
The mass $m_{h}^{2}$ can be derived from the second derivative of 
the effective potential evaluated at the point $\phi = \phi_{0}$, 
\bea
\label{masspot}
m_h^2 = \frac{\partial^2}{\partial \phi^2}
\left\{
-\frac{i}{2} \left( \frac{1}{2\pi \hat{R}} \right) 
\sum_n \int \frac{d^4k}{(2\pi)^4}~{\rm ln}
\left[-k_\mu k^\mu + \left( \frac{n}{\hat{R}} \right)^2 \right]
\right\} |_{\phi = \phi_{0}}.  
\eea
It is easy to see that the Higgs mass vanishes 
in the $\hat{R} \to \infty$ limit as we expect:   
\be
\label{infty}
m_h^2 \to \frac{\partial^2}{\partial \phi^2}
\left\{
-\frac{i}{2}\int \frac{d^4k dk_y}{(2\pi)^5}{\rm ln}[-k_\mu k^\mu + k_y^2]
\right\} |_{\phi = \phi_{0}} = 0, 
\ee
where 
\be
\frac{1}{2\pi \hat{R}}\sum_n \to \int \frac{dk_y}{2\pi}
\ee
is understood. The finite $m_h^2$ for the case of finite $\hat{R}$ may be readily 
obtained by performing the derivative (\ref{masspot}): 
\bea
\label{scalarmass2:g}
m^2_h &=& \left( -\frac{i}{2} \right) \left( \frac{1}{2\pi \hat{R}_0} \right)
\sum_n \int \frac{d^4k}{(2\pi)^4} \nonumber \\
&& \times \left\{ \frac{1}{4} 
\ln \left[ -k_\mu k^\mu + \left( \frac{n}{\hat{R}_0} \right)^2 \right] 
-2\frac{\left( \frac{n}{\hat{R}_0} \right)^2}
{k_\mu k^\mu - \left( \frac{n}{\hat{R}_0} \right)^2}
- \frac{\left( \frac{n}{\hat{R}_0} \right)^4}
{\left\{ k_\mu k^\mu - \left( \frac{n}{\hat{R}_0} \right)^2 \right\}^2}
\right\},  
\eea 
which just recovers the result (\ref{scalarmass:g}) by explicit calculation of Feynman diagrams. 
Therefore the remaining calculation is the same as in the explicit calculation 
and we get the result for the Higgs mass-squared 
\be
m_H^2 = -\frac{25\zeta(5)}{4\pi^5 M_{pl}^2 \hat{R}_0^4},   
\ee

Although the present calculation is that for the quantum effect due to the scalar loop, 
it is straightforward to extend the obtained result to the case 
in which fields with various spins (such as a graviton or a vector field etc)  
run in the loop. 
The 1-loop effective potential for other fields is obtained 
by simply multiplying the physical degrees of freedom of polarization 
to the potential for a real scalar field \cite{PP}\footnote{The periodic 
boundary condition with respect to $S^1$ is assumed. 
Also, all fields we consider are massless. 
Massive fields lead to more complicated potential \cite{PP}.}; For bosonic fields 
\bea
\label{otherspin}
V_{{\rm eff}}^{({\rm graviton})}(\phi) &=& 
5V_{{\rm eff}}^{({\rm scalar})}(\phi), \\
V_{{\rm eff}}^{({\rm vector})}(\phi) &=& 
3V_{{\rm eff}}^{({\rm scalar})}(\phi),  
\eea
and for fermionic fields, including a super-partner,  
\bea 
V_{{\rm eff}}^{({\rm gravitino})}(\phi) &=& 
-8 V_{{\rm eff}}^{({\rm scalar})}(\phi),  \\ 
V_{{\rm eff}}^{({\rm fermion})}(\phi) &=& 
-4 V_{{\rm eff}}^{({\rm scalar})}(\phi). 
\eea 
Hence, a general formula for $m_{H}^{2}$ can be written in terms of 
the numbers of each type of field, $n_{{\rm graviton}}$ etc.: 
\be 
m_{H}^{2} = - \left\{  
(5 n_{{\rm graviton}} + 3 n_{{\rm vector}} + n_{{\rm scalar}}) - 
(8 n_{{\rm gravitino}} + 4 n_{{\rm fermion}})
\right\} 
\cdot \frac{25\zeta(5)}{4\pi^5 M_{pl}^2 \hat{R}_0^4}.  
\ee

Before closing this section, 
we comment on the issue of the radius stabilization. 
The resulting finite Higgs mass squared (\ref{higgsmass}) is negative, 
which implies the instability of our simplified model. 
In fact, the form of the effective potential indicates that 
the radius shrinks to zero (``Casimir effect" \cite{AC}).  
To avoid this situation, we need some extension. 
We first note that adding massless fields does not stabilize the radius 
since the form of the effective potential does not change 
and only the overall coefficient, namely the number of the degrees of 
freedom of physical polarizations, 
is different. If the overall sign of the effective potential is positive, 
the radius goes to infinity. 
On the other hand, if the overall sign is negative, 
the radius shrinks to zero. 
One of the ways to realize the radius stabilization is to introduce 
a massive bulk scalar field. 
If the scalar mass is denoted as $m$, 
the potential minimization tells us that the radius becomes roughly 
${\cal O}(m^{-1})$. 
Even if massive fields are included, the finiteness of 
the Higgs mass remains unchanged.
Thus, we believe that our diagrammatic analysis to clarify the finiteness 
still holds.

\section{Summary and concluding remarks} 
Motivated by the Gauge-Higgs unification scenario, 
we have considered the possibility to solve the hierarchy problem 
by using the framework of the Gravity-Gauge-Higgs unification scenario. 
Taking a five dimensional gravity theory coupled with a bulk scalar field 
as a prototype model, we have explicitly calculated 1-loop corrections 
to the mass-squared of the Higgs originating 
from the extra space component of the metric tensor $g_{55}$. 
We have clarified the mechanism for the quadratic divergence to be 
cancelled in a diagrammatic way. It has been argued that to get the calculable 
finite Higgs mass, summing up all K-K modes running in the loop diagram is crucial, in order to 
maintain the general coordinate transformation invariance.   
In such calculation, the bubble diagram and the tadpole diagram contributions are important 
to obtain a finite mass, although these diagrams naively seem to 
have no contributions to the Higgs mass-squared $m_{H}^{2}$.  

The calculation was performed by two different ways, i.e. by direct calculation of 
Feynman diagrams and by utilizing the effective potential. A detailed calculation was made 
for the quantum correction due to the bulk scalar field. The obtained result has been 
generalized to the contributions from a variety of fields with various spins, including 
graviton itself, and a general formula has been obtained for the Higgs mass-squared.   

Having shown that the mechanism to solve the hierarchy problem (to realize the stability of 
$m_{H}$) works, the next step will be to device a realistic theory as the theory of elementary 
particles. Then, the issues we can immediately think of are the following. 

First, the gauge symmetry of the theory, with $A_{\mu}$ being its gauge boson, should be U(1), as in 
the original K-K theory. We obviously need to extend the gauge symmetry to non-Abelian symmetries, 
in order to incorporate the $SU(2) \times U(1)$ of the standard model. If we consider a 
7D gravity theory compactified 
on $S^2 \times S^1$, an $SU(2) \times U(1)$ gauge group arises as the isometry of 
$S^2 \times S^1$ \cite{Witten}.  

Secondly, if the extra space components of the metric tensor are to play the role of Higgs 
fields, they should be responsible for the spontaneous gauge symmetry breaking (SSB). 
In the prototype model we discussed in this paper the gauge symmetry is $U(1)$ and 
SSB is not possible. However, once the extra space is enlarged, the extra space 
components of the metric generally belong to some non-trivial representations of 
the non-Abelian gauge group (isometry), and SSB is possible, in a similar manner with that in 
the Hosotani mechanism \cite{Hosotani} in the Gauge-Higgs unification scenario. More intuitively, one 
may say that if the compactified space is deformed from a symmetric space, such as $S^{n}$, 
in the ground state, the SSB is realized, as explicitly shown in \cite{L}.   
 
Next issue is that the obtained finite $m_{H}$ (\ref{higgsmass}) should be comparable to 
the weak scale $M_{W}$ for the complete solution of the hierarchy problem. One interesting 
possibility to realize it might be to make $\hat{R}_{0}$ in (\ref{higgsmass}) of the order of 
``intermediate scale", $1/\hat{R}_{0} \sim \sqrt{M_{pl} M_{W}} \sim 10^{11} \ ({\rm GeV})$ to 
realize $m_{H} \sim M_{W}$. We, however, note that if it is the case the gauge coupling of $A_{\mu}$ 
to matter fields may be ${\cal O}(\frac{1}{M_{pl} \hat{R}_{0}}) = {\cal O}(10^{-8})$, which 
is too small to account for the magnitude of $e = \sqrt{4\pi \alpha}$. A possible breakthrough may be to invoke to simply-connected extra space, such as $S^{n} \ (n > 1)$, 
in which A-B type effect seems to be irrelevant. As a first step, it might be interesting to consider 
a 6D gravity theory compactified on $S^2$. In fact, in the Gauge-Higgs unification scenario \cite{HIL}, 
1-loop correction to Higgs mass has been calculated in scalar QED theory compactified on $S^2$, 
where the mass was found to vanish identically. This is because any loop on $S^2$ can shrink 
to a point, then the non-trivial Wison loop, which is the non-local operator to yield finite $m_{H}$, 
is not allowed. Thus, it deserves to investigate whether this fact also applies to 
the Gravity-Gauge-Higgs unification scenario. In this way, we may be able to realize a small 
$m_{H}$ even for very small extra dimension of the size of $1/M_{pl}$. If this mechanism works 
it will be desirable to study the Higgs mass in a 7D gravity theory compactified on $S^2 \times S^1$. 

Finally, in the Gauge-Higgs unification scenario, 
the finite mass $m_{H}$ can be understood as due to the appearance of 
a non-trivial Wison loop. However, it is unclear for us whether a similar 
understanding also holds for the present Gravity-Gauge-Higgs unification scenario. 
A naive correspondence of the Wilson loop seems to be a line integral of 
Christoffel symbols, not $g_{55}$ itself, whose physical meaning has not been clarified 
so far.  

These issues are left for future investigations.

%
%
%
\subsection*{Acknowledgment}

We would like to thank T. Inami for useful informative discussion. 
The authors would like to acknowledge the grant meeting ``Gauge-Higgs Unification and Extra Dimension" 
held at Kobe University, Japan (Jan., 2004), where this work started.  
The work of C.S.L. was supported in part by the Grant-in-Aid for Scientific Research 
of the Ministry of Education, Science and Culture, No.15340078.  
N.M. is supported by Special Postdoctoral Researchers Program at RIKEN. 

\end{document}